# A New Model of Array Grammar for generating Connected Patterns on an Image Neighborhood

G. Vishnu Murthy*‡, Pavan Kumar C$, Dr. Vakulabharanam Vijaya Kumar#

*HOD – CSE, Anurag Group of Institutions & Research Scholar, JNTU, Hyderabad

‡Corresponding Author; Email: gvm189@gmail.com

$Student, M. Tech.– CSE, Anurag Group of Institutions Email:pforpavank@gmail.com

#Professor and Dean – Computer Sciences, Anurag Group of Institutions & Head, Center for Advanced Computational Research,

Email:vakulabharanam@gmail.com

**Abstract:** *Study of patterns on images is recognized as an important step in characterization and classification of image. The ability to efficiently analyze and describe image patterns is thus of fundamental importance. The study of syntactic methods of describing pictures has been of interest for researchers. Array Grammars can be used to represent and recognize connected patterns. In any image the patterns are recognized using connected patterns. It is very difficult to represent all connected patterns (CP) even on a small 3 x 3 neighborhood in a pictorial way. The present paper proposes the model of array grammar capable of generating any kind of simple or complex pattern and derivation of connected pattern in an image neighborhood using the proposed grammar is discussed.*

## 1. INTRODUCTION

Today string grammars are studied widely in the field of computer science, mathematics and linguistics since they describe various forms of language constructs. The string grammar plays a significant and crucial role in the analysis of any language especially in high level languages. Similarly the study of syntactic methods of describing pictures considered as connected, digitized finite arrays in a two-dimensional plane[1] have been of great interest for many researchers. There are two different types of models one, puzzle languages [2] and the other, recognizable picture languages [1]. The former introduced to solve certain problems of tiling, is a type of Rosenfeld model [3]. In the context free case, the generative capacity of puzzle grammars is the same as that of Context Free array grammars [2] but in the case of basic puzzle grammars consisting of an extension of the right linear rules, the generative power is higher than that of regular array grammars [5] . The second model was introduced in an attempt to extend the notion of recognizability in one dimension to two dimensions. In the one-dimensional case, the notions of languages generated by the right linear (left linear) grammars, languages accepted by finite automata (deterministic or non-deterministic), rational languages, recognizable languages all coincide. The new model of recognizable picture language extends to two dimensions the characterization of the one-dimensional recognizable languages in terms of the alphabetic morphism of local languages. This has good closure properties but is still not closed under complementation and has an undecidable emptiness problem. The class has been shown to be equivalent to the class recognized by on-line tessellation automaton [6], a kind of cellular automata and has also been characterized in terms of existential, monadic, second-order definable picture languages [7]. In 1990 two different types of picture language models called Puzzle languages and recognizable picture languages describing digitized pictures using 2 – D plane are introduced [2, 3, 4, 8]. G Siromoney et. al[9] derived parallel/sequential model capable of generating interesting classes of pictures that do not maintain a fixed proportion. Later KAG is motivated by the intricate patterns and found in the kolam (the folk are of India, also called rangoli or rangavalli) and the need to have array rewriting rules. Motivated by this[8] a second set of rectangular models (hereafter called rectangular kolam arrays) which are more powerful and are suited to generating patterns that maintain fixed proportion. In order to show growth along the edges and to extend Lindenmayer systems [12] to arrays, G Siromoney et. al. introduced rectangular and radial L-models [10,11]. These models compare favorably with kolam models in their generative power, but they are easier to operate.

The present paper is organized as follows. The section two deals with Role of patterns in image processing. In section three the definition of Isometric Array Grammar is given. The section four discusses the need and definition of connected pattern array grammar (CPAG). The section five presents examples of derived CP by the proposed CPAG. The section six gives the conclusions.

## 2. Role of Pattern in Image Processing

Today images are obtained without noise with high resolution, accuracy, without any illumination and reflection effects after the development of powerful high resolution digital cameras, powerful scanners and other image capturing devices. That's why Image analysis and characterization was widely studied over the last three decades in a variety of applications, including medical imaging, pattern recognition, industrial inspection, age classification, face recognition, texture classification, and texture based image retrieval.

Patterns have been recognized as an important attributes of image data. They are used extensively in the visual interpretation of image data, in which pattern is often more important than the other image attributes. A simple pattern of a neighborhood can be considered as one of the image primitive feature. Image patterns can often be used to recognize familiar objects in an image or retrieve images with similar pattern. Study of patterns plays a significant role in texture analysis.

The present paper assumes image texture as characterized not only by the grey value at a given pixel, but also by the grey value pattern in a neighborhood surrounding the pixel. The ability to efficiently analyze and describe textured patterns is thus of fundamental importance. A simple pattern of a neighborhood can be considered as one of the texture primitive feature. Textural patterns can often be used to recognize familiar objects in an image or retrieve images with similar texture from a database.

## 2.1 Representation of connected patterns on a 3 x 3 neighborhood

Though the study and derivation of patterns plays a significant and crucial role in image texture analysis, most of the researchers find it difficult to represent/denote various patterns in a neighborhood. For example, a 3 * 3 neighborhood will have nine pixels and representing CP in a pictorial way is very difficult. The proposed paper derives an array grammar model that efficiently represents various connected patterns in the neighborhood.

Based on the above significance of patterns and their role in various image analysis, classification, recognition and representation issues, few researchers concentrated on developing a Grammatical representations to patterns. These representations are called as Array Grammars (AG).

If we count "Connected Patterns"(CP) in a 3 x 3 neighborhood by always assuming the central pixel of the window is a grain or one. In the following figures '0' indicates no grain and '1' indicates a grain. There will be eight different formations of CP with two grain components by fixing one of the grains at pixel location (0,0) on a 3×3 mask as shown in Fig.1.

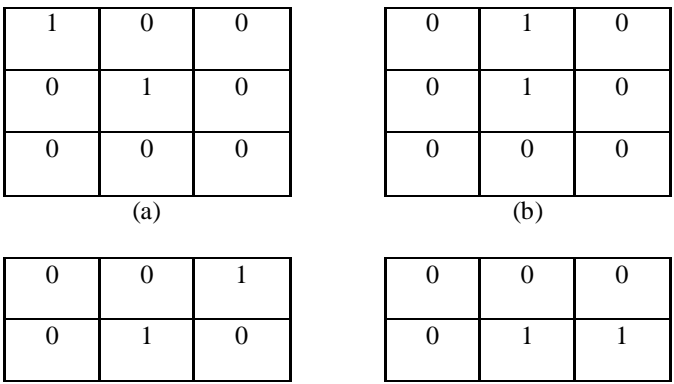

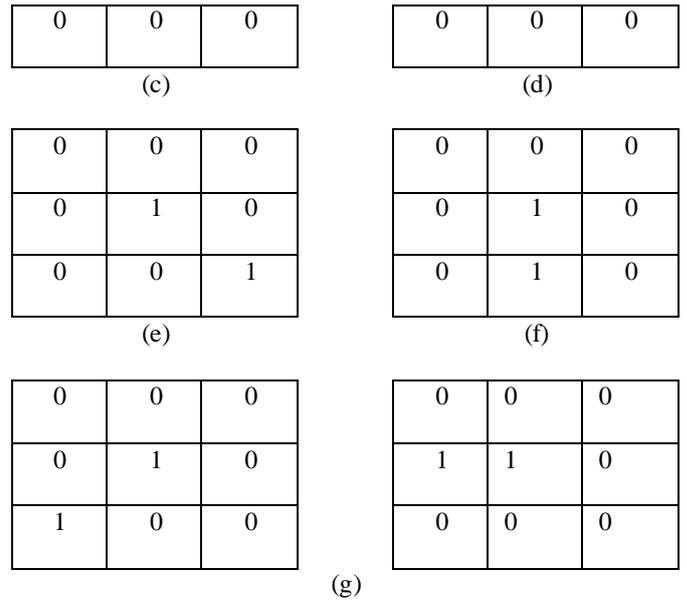

Fig.1 Representation of CP with two grain components by fixing one of the grain component at (0, 0)

In the similar way there will be twenty eight formations of CP with two grain by positioning one of the grains at the pixel location (0, 0) and changing the position of the other without repetition of patterns. Proceeding in the same manner by increasing the fixed grain components, many connected patterns can be generated on a 3×3 window. This indicates that we can generate different patterns by fixing grains. Representing such a huge number of patterns is tedious task. The present paper addresses this, by deriving a new array grammar called "Connected Patten Array Grammar (CPAG)".

## 3 Array Grammars

The language of an array grammar has been defined as the set of finite, connected terminal arrays. The terminal array components are continuous forming connected patterns. To ensure connectedness of patterns few AG models have made use of special '#' symbol surrounding the start symbol and also they play an extensive role in maintaining the geometrically isometric properties of production rules[13].

The digitized finite arrays in a two-dimensional plane have been of great interest for many researchers. Picture languages generated by array grammars or recognized by array automata have been advocated since the 1970s for problems arising in the framework of pattern recognition and image processing [14]. The language of an array grammar has been defined by many researchers as the set of finite connected terminal arrays, surrounded by #'s, that can be derived by an initial/starting symbol S surrounded by #'s[13].

Isometric (or Isotonic) AG introduced by Rosenfeld[3] is a formal model of two dimensional pattern generation.

Definition: An isometric array grammar (IAG) is a system defined by

$$G = (N, T, P, S, \#)$$

where N is a finite nonempty set of nonterminal symbols, T is a finite nonempty set of terminal symbols (N ∩ T = Φ), S (є N) is a start symbol, # (doesn't belong to N U T) is a special blank symbol.

P is a finite set of rewriting rules of the form α → β where α and β are words over N U T U {#} and satisfy the following conditions:

1. The shapes of α and β are geometrically identical.
2. α contains at least one nonterminal symbol.
3. Terminal symbols in α are not rewritten by the rule α → β.
4. The application of the rule α → β preserves the connectivity of the host array

## 4 Generation of Connected Patterns

Study of patterns on textures is recognized as an important step in characterization and classification of texture. Various approaches are existing to investigate the textural and spatial structural characteristics of image data, including measures of texture [16], Fourier analysis [15,17], fractal dimension [18], variograms [19,20,21,23] and local variance measures [22].

The present paper after studying the significance of pattern representation and also to avoid and to overcome the tedious process of representing CP as described in section 2.1, derived a new array grammar called Connected pattern Array Grammar (CPAG). The CPAG is a useful representation to derive CP on any neighborhood. No researcher has concentrated in the grammars for deriving connected patterns that can overcome the tedious way of pictorial representation.

A CPAG is defined by 5 tuple form where G = ({S, A}, {a, b}, P, S, #) where S and A are the non-terminals, a and b are terminals, P is the set of production rules as mentioned below, S is the start symbol and # is the special blank symbol.

The set P of rules is as follows:

$$S\# \rightarrow aA \mid bA$$

$$\frac{S}{\#} \rightarrow \frac{a}{A} \mid \frac{A}{a} \mid \frac{b}{A} \mid \frac{A}{b}$$

$$A\# \rightarrow aA \mid bA \mid \#A$$

$$\frac{A}{\#} \rightarrow \frac{a}{A} \mid \frac{A}{a} \mid \frac{b}{A} \mid \frac{A}{b} \mid \frac{\#}{A}$$

$$\frac{\#}{A} \rightarrow \frac{A}{a} \mid \frac{A}{b}$$

$$\#A \rightarrow Aa \mid A\# \mid aA \mid bA \mid Ab$$

$$A \rightarrow a \mid b$$

In the above representation, the terminal symbol a is treated as grain component or with intensity value 1, b is represented as non-grain component or 0 intensity value.

The proposed CPAG is an Isometric Array Grammar as each pair of rules in G satisfies the condition of Isometric Array Grammar production rules and also the production rules satisfies the condition of Context Free Array Grammar as α contains exactly one non terminal and possibly one or more #s.

## 5 Generation of CP by the proposed CPAG

The following productions of the proposed CPAG derives CP with one grain component (central pixel is assumed to be a grain) on a 3x 3 neighborhood.

| S | # | # |   | b | A | # |   | b | b | A |   |
|---|---|---|---|---|---|---|---|---|---|---|---|
| # | # | # | → | # | # | # | → | # | # | # | → |
| # | # | # |   | # | # | # |   | # | # | # |   |

| b | b | a |   | b | b | a |   | b | b | a |   |
|---|---|---|---|---|---|---|---|---|---|---|---|
| # | # | A | → | # | A | b | → | A | a | b | → |
| # | # | # |   | # | # | # |   | # | # | # |   |

| b | b | a |   | b | b | a |   | b | b | a |   |
|---|---|---|---|---|---|---|---|---|---|---|---|
| b | a | b | → | b | a | b | → | b | a | b | → |
| A | # | # |   | b | A | # |   | b | b | A |   |

| b | b | a |
|---|---|---|
| b | a | b |
| b | b | b |

## 6 Conclusions

The proposed CPAG attempted and overcome the tedious way of representing the connected patterns in an image neighborhood. Since the same production rules are used to generate any kind of connected pattern no necessity of defining new set of production rules for each kind of pattern in CPAG. The proposed CPAG is easy to understand as it is more or less similar to Context Free class of string grammars and also easy to derive any kind of connected patterns as it is having less number of production rules. The proposed work can be extended for deriving patterns of signal processing. Also, a recognizing device may be constructed to recognize CPAG derived patterns.

**References:**

[1] D. Giammarresi, A Rivestivo, "Recognizable picture languages", Int. Journal of Pattern Recognition, Artificial Intell. 6(1992) 241

[2] M. Nivat, A Saoudi, K G Subramanian, R Siromoney, V R Dare, "Puzzle Grammars and Context free array grammars", International Journal of Pattern Recognition Artificial Intelligence, 5 (1991) 663

[3] A. Rosenfeld, "Picture Languages (formal models for picture recognition)", Academic Press, New York 1979

[4] R. Siromoney, "Array Languages and Lindermayer systems" – a survey in the book of L. G. Rozenberg, A Salomaa (Eds.) Springer, Berlin, 1985

[5] K G Subrmanian, R Siromoney, V R Dare, A Saoudi, "Basic Puzzle languages", Int. Journal of Pattern Recognition, Artificial Intell. 9(1995) 763

[6] K. Inoue, I. Takanami, "A characterization of recognizable picture languages", Lecture notes in coputer science, Vol.654, Springer, 1993

[7] D. Giammarresi, A Rivestivo, S. Seibest, W Thomas, "Monadic second order logic over rectangular pictures and recognizability by tiling systems", Inform. Comput. 125 (1996) 32

[8] G Siromoney, R Siromoney, K Krithivasan, "Picture Languages with array rewriting rules", Inform. Control 22 (1973) 447

[9] G Siromoney, R Siromoney and K Krithivasan, "Abstract families of matrices and picture languages", Computer Graphics and Image Processing, 1, 1972, 284, 307

[10] G Siromoney, R Siromoney, "Radial Grammars and radial L systems", Computer Graphics and Image Processing, 4, 1975, 361-374

[11] R Siromoney and GSiromoney, "Extended Controlled table arrays", TR-304, Computer Science Center, University of Maryland, May, 1974.

[12] A Lindenmayaer, "Mathematical models for cellular interactions in development", I and II, J. Theoretical Biol, 18, 1968, 280-315

[13] A Rosenfeld, "Array Grammar Normal Forms", Information and Control, 23, 173-182, 1973

[14] R Siromoney, K G Subramanian, V R Dare, D G Thomas, "Some results on Picture Languages", Pattern Recogniton 32, 295-304, 1999

[15] Moody A. and Johnson D. M., "Land-surface phenologies from AVHRR using the discrete Fourier transform", Remote Sens. Environ., Vol. 75, pp. 305-323, 2001.

[16] Richards J. A. and Xiuping J., "Remote Sensing Digital Analysis: An Introduction. Germany: Springer-Verlag, Vol.3, pp.363-363, 1999.

[17] Zhang M., Carder K., Muller-Karger F. E., Lee Z. and Goldgof D. B. "Noise reduction and atmospheric correction for coastal applications of landsat thematic mapper imagery", Remote Sens. Environ., Vol. 70, pp. 167-180, 1999.

[18] Burrough P. A., "Multiscale sources of spatial variation in soil, the application of fractal concepts to nested levels of soil variation," J. Soil Sci., Vol. 34, pp. 577–597, 1983.

[19] Atkinson P. M. and Lewis P., "Geostatistical classification for remote sensing: An introduction", Comput. Geosci., Vol. 26, pp. 361–371, 2000.

[20] Curran P. J., "The Semi-variogram in Remote Sensing: An Introduction", Remote Sens. Environ., Vol. 24, pp. 493–507, 1988.

[21] Treitz P,"Variogram analysis of high spatial resolution remote sensing data: An examination of boreal forest ecosystems", Int. J. Remote Sens., Vol. 22, pp. 3895-3900, 2001.

[22] Woodcock C. E. and Strahler A. H., "The factor of scale in remote sensing", Remote Sens. Environ., Vol. 21, pp. 311-332, 1987.

[23] Woodcock C. E., Strahler A. H. and Jupp D. L. B. "The use of variograms in remote sensing II: Real digital images", Remote Sens. Environ., Vol. 25, pp. 349-379, 1988.